\documentclass[onecolumn]{article}

\usepackage[dvips]{graphicx}
\usepackage{amssymb,amsfonts,amsmath}

\date{}

\newtheorem{remark}{Remark}
\newtheorem{definition}{Definition}
\newtheorem{problem}{Problem}
\newtheorem{proposition}{Proposition}

\begin{document}

\title{Geometric Path Integrals. A Language for Multiscale Biology and Systems Robustness. \bigskip
\\
{\large \it This paper is dedicated to the memory of\\ Leon Ehrenpreis 1930-2010}}

\author{D. Napoletani \thanks{Center for Applied Proteomics and Molecular Medicine,
George Mason University, Manassas, VA 20110 USA, and Schmid College of Science, Chapman University, Orange, CA 92866. Email: dnapolet@gmail.com}, E. Petricoin
\thanks{Center for Applied Proteomics and Molecular Medicine,
George Mason University, Manassas, VA 20110 USA.}, D. C. Struppa \thanks{Department of Mathematics and Computer Science, Chapman University, Orange, CA 92866}.}

\maketitle

\begin{abstract}

In this paper we suggest that, under suitable conditions, supervised learning can provide the basis to formulate at the microscopic level quantitative questions on the phenotype structure of multicellular organisms.
The problem of explaining the robustness of the phenotype structure is rephrased as a real geometrical problem on a fixed domain. We further suggest a generalization of path integrals that reduces the problem of deciding whether a given molecular network can generate specific phenotypes to a numerical property of a robustness function with complex output, for which we give heuristic justification. Finally, we use our formalism to interpret a pointedly quantitative developmental biology problem on the allowed number of pairs of legs in centipedes.

\end{abstract}

\section{Introduction}

Leon Ehrenpreis was a singular mathematician. Not only he had a gift and a vision for a deep understanding of mathematics, but he had a passion for the construction of overarching approaches, that would allow a general comprehension of vast areas of mathematics. This passion is embodied in his two masterpieces, Fourier Analysis in Several Complex Variables, \cite{ehrenpreis1}, and The Universality of the Radon Transform, \cite{ehrenpreis2}, but is also apparent in the many papers he published, for example, on wide generalizations of the Edge-of –the Wedge theorem.

Reading Ehrenpreis' works, we are reminded of a beautiful phrase that Kawai, Kashiwara, and Kimura insert in \cite{kkk}, just after the proof of the Watermelon Theorem:``It was like looking down at the valley after reaching the peak of the hill''.

This aspiration to a global view of mathematics, that would offer insights, even in advance of a fully realized technical  description of that view, is part of what made Ehrenpreis' work uniquely captivating, and uniquely fertile. Many mathematicians, of at least three generations, have worked to understand, formalize, explain, refute, demonstrate, statements in Ehrenpreis' work. It is for this reason, that Ehrenpreis' student, C.A.Berenstein, once noted in his review \cite{carlos} of \cite{ehrenpreis2} that ``a book that is worth studying, although mining may be a more appropriate word, as the reader may find the clues to the keys he's searching for to open up subjects that are seemingly unrelated to this book. Thus, one finds at the end that the title is justified.''.

It is in this spirit, but with a deep sense of humility and with a full awareness of our limitations, that we would like to present, in this paper, a proposal, a strategy for a way to mathematically understand and describe one of the fundamental problems (the fundamental problem?) of modern biology: how can we understand macroscopic biological traits from our knowledge of molecular level information.

The proposal we make is inspired by the instrument of path integrals, which is probably the most enduring legacy of Richard Feynman \cite{feynman}, and for which we suggest here a tentative generalization to provide a plausible tool for the description of the macroscopic properties of a biological system.

Our specific point of departure lies in an important, and somewhat surprising, fact.
In \cite{minelli} chapter 9 it is reported that, out of all known species of centipedes, there are about 1000 species with 15 pairs of legs, none with 17 or 19 pairs, several with 21 and 23 pairs, and a few distributed over a very large range from 27 to 191. No centipedes have an even number of pairs of legs, and some species have a stable interspecies number of pairs of legs, while some others display a variability of the number of legs among individuals.

Can we explain in any quantitative way this striking pattern of gaps with respect to the dynamics at fine molecular scale? And how do we express the remarkable robustness of the resulting phenotypes? To be more specific, we state the following
\begin{problem} [Centipedes' Segmentation (CS) Problem] Show that it is impossible to have centipedes with an even number of pairs of legs, or with 17 or 19 pairs of legs.
\end{problem}

Clearly if we can suitably quantify this problem, we will be able to generalize the question to the full gap structure that we have described in the previous paragraph. Moreover, we take the CS problem simply as emblematic of the variety of developmental biology problems in which we observe strong constraints on the phenotype (\cite{minelli}, page 86), without having a conceptual frame through which to approach these problems.

In general, molecular biology approaches to developmental biology problems such as the CS problem require the construction of an appropriate map that can relate microscopic variables with macroscopic outputs. But, since a mechanistic way to relate these variables is often absent, and it is indeed problematic even to quantify phenotype properties \cite{phenomics}, it is challenging to even express these questions in a proper mathematical setting. Even though we are aware of the dangers of suggesting a general frame that is not fully developed, we still believe (in view of the difficulty of the problems at hand) that it is worthwhile to attempt to set up a language and a set of techniques through which these problems might be approached.

Ideally, we would need to study entire sets of models at once, since we would expect the micro/macro maps to be stable under wide variations of the parameters in the model of the microscopic dynamics. At the same time, such tremendous model variability should be understandable in a compact, possibly analytical setting to have any hope of providing computationally feasible answers.

In this paper we suggest that supervised learning and the resulting classification functions \cite{hastie} can provide the basis to formulate at the microscopic level questions on the phenotype, such as the centipedes' segmentation problem, provided that the classification function satisfies some suitable growth conditions. The problem of phenotype robustness can then be rephrased as a problem of real geometry on a fixed domain. General methods to solve such problems are still in their infancy \cite{RAG} and we propose a generalization of path integrals that allows to reduce the problem of class belonging and of phenotype robustness to  specific questions on functions with complex output. Finally, we will show how to reduce the CS problem to a problem on the global properties of these functions. Other problems about the restricted variability of phenotypes in developmental biology could be formulated in similar ways.

\section{A Geometrical Robustness Condition}

Let $X=(X_1,\dots,X_N)$ be the activation level of a set of proteins (genes, metabolites, or combinations of) at some time $t_0$ and assume that we have access to the derivatives $\dot X=(\dot X_1,\dots,\dot X_N)$ of those levels at the same time $t_0$ (in practice this means that we measure the proteins at two very close time points); and suppose the biological samples from which $(X,\dot X)$ is measured can be classified in a set of $M$ classes $C_1,\dots,C_M$.

We assume the protein measurements are taken at the embryo stage of development of the individuals in the CS problem. Clearly these measurements will be some sort of average of the activity levels of several cells \cite{RPPA}, even though single cell measurements can be envisioned \cite{singlecell}. We further assume that the underlying network of interactions is stable (i.e. no essential parameter variation that changes significantly the dynamics) over a short time range. Without excluding in principle other possible state variables, we focus our attention on protein networks as these are believed to be evolutionary more stable \cite{refMinelli1}.

Call $B(X,\dot X)$ an instance of the biological class associated to $(X,\dot X)$ (in the case of the CS problem, this will be the number of pairs of legs of an adult individual centipede). We assume that we have access to a training set of instances for each class, so that we can build a special type of classifier that has the following structure:

\begin{definition}[Interval Classifier] A function $F(X,\dot X)$ is an interval classifier for the classification problem with classes $S_m$, $m=1,\dots,M$, if it satisfies: $m-1<F(X,\dot X)<m$ when $B(X,\dot X)\in C_m$ $m=1,\dots,M$.
\end{definition}

Though not strictly necessary at this point, it is useful for our subsequent analysis of Propositions 1 and 2 to require that $F$ is bounded at infinity and analytic, and therefore we assume for simplicity that the classifier function $F$ is a neural network with exponential sigmoidal activation functions (\cite{classbook2}, chapter 10).

\begin{remark}
At a fundamental level, it is not necessarily possible to identify a subset of molecular variables that is indeed predictive for the phenotype characteristic that we are interested in. This problem is not exclusive to our setting, but it is a major difficulty in all approaches that try to bridge molecular biology with the study of phenotype characteristics. The very existence of an accurate classifier $F$ depends on the identification of such variables. Note that our setting requires to estimate not only state variables, but their derivatives as well. In classical rational mechanics state variables and their first derivative are sufficient to characterize a system for all future times (in a variational, Lagrangian setting \cite{arnold}). While first derivatives are also sufficient for our formal analysis of phenotype classification problems, it would be interesting to understand how many derivatives are truly necessary to have effective classifiers for these types of problems.
\end{remark}

Note that the way we define the multi-classes classifier $F$ is not the standard one, in which an $M$-classes problem is usually approached by having a vector of $M$ output classification functions (\cite{classbook2}, page 331). For reasons that will be clear when we reinterpret analytically the CS problem in Problem 2 of the last section, we not only use a single function $F$ for the multi-class problem, but we also require that all instances belonging to a certain class must be within a given interval. This superimposes a stronger metrical structure on the classification problem.

\begin{remark}
In the context of neural networks, the request of interval classifier for training instances within each class transforms the unconstrained optimization problem usually associated to finding the classifier function (\cite{classbook2}, page 335) into a constrained optimization problem.
\end{remark}

Take now a slack variable $Y$ and consider the function  $F(X,\dot X)-Y$, with $(X,\dot X)\in D\times \dot D$, where $D$ is the set of biologically meaningful conditions for $X$ and $\dot D$ is the set of biologically meaningful conditions for $\dot X$. Then if $B(X,\dot X)\in C_1$ there exists $Y$ with $0<Y<1$ such that $F(X,\dot X)-Y=0$, so that the condition $B(X,\dot X)\in C_1$ can be rewritten as  $F(X,\dot X)-Y=0$, $(X,\dot X)\in D\times \dot D$, $0<Y<1$. Similarly, $B(X,\dot X)\in C_m$ becomes $F(X,\dot X)-Y=0$, $(X,\dot X)\in D\times \dot D$, $m-1<Y<m$.

It is reasonable to suppose that $X$ is in fact a state variable of an ordinary differential equation (ODE) network $\dot x=f(x,a_0)$, $x=(x_1,\dots,x_N)$, $f=(f_1,\dots,f_N)$, $f$ a vector of polynomials in $x$, modeling ODEs with polynomials or power functions has proven itself to be very flexible for systems of molecular reactions \cite{voit}. We further ask that $f(x,a)$ is an analytic function in $a$. As the condition of analytic structure of the classifier $F$ itself, the analyticity of $f(x,a)$ in $a$ will be important in the justification of Proposition 2.

A network of biological significance will usually depend on a large number of parameters that will depend on the environment where the variable of the network actually act and live \cite{guna}, this is the reason we allow a dependance from the parameter vector $a$ in the ODE network.
Write the dependence of $\dot x$ from $f(x,a)$ explicitly in $F(X,\dot X)-Y=0$, i.e., $F(X,f(X,a))-Y=0$. The condition of class belonging can be written as:

\begin{definition}[Network Classification] A network $\dot x=f(x,a_0)$ generates phenotypes belonging to class $C_m$ if
\begin{equation}
\label{class1}
\exists X\in D,\,m-1<Y<m\,:\,\,F(X,f(X,a_0))-Y=0.
\end{equation}
\end{definition}

Equation \eqref{class1} is a condition for the network $\dot x=f(x,a_0)$ to give rise to states that belong to one of the classes we are considering.
Note that the domain $D$ of $X$ constrains the domain $\dot D$ through the relation $\dot D=f(D)$.

The macroscopic phenotypic states of an organism are believed to be robust under wide ranging changes of parameters \cite{guna}. Therefore, for a realistic network, \eqref{class1} should be satisfied for all parameters in a region A, where $A$ is some sizable neighborhood around a nominal value $a=a_0$ of the parameter $a$. In other words:

\begin{definition}[Class Robustness]
 A phenotype class $C_m$ is robust if the zeros of the function $F(X,f(X,a))-Y=0$ are persistent in a region $A$ of parameters, i.e.:
\begin{equation}
\label{classA}
\forall a\in A,\,\exists X\in D,\,m-1<Y<m\,: F(X,f(X,a))-Y=0.
\end{equation}
\end{definition}

\begin{remark}
We assume that, for each $a$, $\dot x=f(x,a)$ is capable of generating $(x,\dot x)$ belonging to a single class $C_m$, to avoid, in the CS problem, the paradoxical situation in which the predicted number of pairs of legs can change in a given centipede with time. We assume instead that the embryo is committed to its specific segmentation within a large time frame where we could measure our state variables.
\end{remark}

\section{Stable Zeros and Path Integrals}

In the previous section we described a condition \eqref{classA} that must be satisfied if $\dot x=f(x,a)$ is to generate robustly states that belong to a class $C_m$. The problem with this condition is that it requires identification of zeros of a (non-algebraic) function over a real domain, and moreover it requires us to establish that these zeros are stable under a wide variation of parameters. This is problematic as it is difficult to establish the existence of solutions of real equations on domains, even in the algebraic case (\cite{RAG2}).

In this section we show that a generalized path integral \cite{feynman} can be built in such a way that a specific condition on this integral corresponds to the verification or falsification of \eqref{classA} over a domain. Path integrals have the remarkable property of giving information, in a single analytical object, about global, collective properties of physical systems, a point of view especially stressed in condensed matter field theory literature \cite{condensed}, and it is this ability that we will try to mirror in the setting of network analysis. We start by building a path integral that is related to \eqref{class1}. Essentially, we will build a domain $G$ and a function $L$ such that if \eqref{class1} is satisfied, then there is at least a path connecting $2$ points in $G$. This path (and a small tubular neighborhood thereof, with squeezed endpoints) will dominate the path integral that we are building and it will allow us to make qualitative conclusions on the value of the integral when \eqref{class1} is verified. We mirror then this analysis for \eqref{classA}. We first go through the technical building of the path integral, before we explain its heuristic interpretation.

The condition $x=(x_1,\dots,x_N)\in D$ in \eqref{class1} can be explicitly written as a condition on each variable, i.e. $d_{nb}<x_n<d_{nt}$, where $d_{nb}$ and $d_{nt}$ are lower and upper bounds on the biologically meaningful values that variable $x_n$ can assume; in principle these values can be measured over repeated in vitro experiments. We can always change variables $x_n \rightarrow \tilde x_n$ so that $-1<\tilde x_n<1$. This is accomplished by setting:

\begin{equation}
\tilde x_n=\frac{2}{d_{nt}-d_{nb}}(x_n-d_{nb})-1 \nonumber
\end{equation}

Similarly, we can force $-1<\tilde y<1$ by setting $\tilde y=2(y-(m-1))-1$. These are invertible linear transforms, so we can write

\begin{equation}
\label{class_tilde}
\exists \tilde x,\tilde y:
\tilde F(\tilde x,\tilde f(\tilde x,a))-(\frac{\tilde y-1}{2}+(m-1))=0,\,\,-1<\tilde x_n<1,\,\,-1<\tilde y<1\,\,a=a_0.
\end{equation}

for some transformed functions $\tilde F$, $\tilde{f}=(\tilde f_1,...,\tilde f_N)$ which are obtained from $F$ and $f$ by replacing $x$ with $\tilde x$ and $y$ with $\tilde y$. We further simplify notation by defining the analytical function

\begin{equation}
\label{H}
 H(\tilde x, \tilde y,a,m)=\tilde F(\tilde x,\tilde f(\tilde x,a))-(\frac{\tilde y-1}{2}+(m-1))
\end{equation}

So we can rewrite condition \eqref{class_tilde}  as:

\begin{equation}
\label{classH}
\exists \tilde x,\tilde y:H(\tilde x,\tilde y, a, m)^2=0,\,on\,\,-1<\tilde x_n<1, \,\,-1<\tilde y<1,\,\, a=a_0.
\end{equation}

We square the function $H$ for purposes that will be clear in the following (see equation \eqref{NPI} and its justification).
The domain restriction on $\tilde x_n$ and $\tilde y$ can also be written as $(\tilde x, \tilde y)\in  \mathcal{S}_1 \times \dots \times\mathcal{S}_1$ where $\mathcal{S}_1$ is the unit interval $[-1,1]$ and we take the cartesian product $N+1$ times. We now introduce a spherical extension of this domain in such a way that on every section of the extended sphere we can formulate a condition similar to \eqref{classH}.

To build the spherical extension, first suppose we work with a single variable $\tilde x_n$, keeping all other variables constant. We embed each point $\tilde x_n$ in the disk $\mathcal{S}_1$ in the space $R^2$ with the map $\tilde x_n\rightarrow (\tilde x_n, 0)$. We want then a basic way to map points $(\tilde x_n,0)$  on $(\mathcal{S}_1,0)$ to points $(\tilde x_{nz}, z)$ in the slices $(*, z)$ for $z$ in $-1<z<1$, and moreover we want the whole set $(\mathcal{S}_1,0)$  to be mapped to the points $(-1,0)$, $(1,0)$ in the limit of $z\rightarrow \pm 1$.

One way to achieve this embedding is through maps $\tilde x_{nz}=\tilde x_n \sqrt{1-z^2}$. Conversely, any point $(\tilde x_{nz}, z)$, in $\mathcal{D}_n=\{-1<z<1,\,\,\,-\sqrt{1-z^2}<\tilde x_{nz}<\sqrt{1-z^2}\}$ can be mapped to a point in $\mathcal{S}_1$ by setting $\tilde x_n=\tilde x_{nz}\frac{1}{\sqrt{1-z^2}}$. If we do a similar mapping for all
$\tilde x_{n}$, and for $\tilde y$ as well, the function $H(\tilde x,\tilde y, a, m)^2$ can be extended to the following function of variables $(\tilde x_{z},\tilde y_z, z, a, m)$:

\begin{equation}
\label{L}
L(\tilde x_z, \tilde y_z, z,a,m)=H(\tilde x_{z}\frac{1}{\sqrt{1-z^2}},\tilde y_z\frac{1}{\sqrt{1-z^2}},a,m)^2
\end{equation}

where $\tilde x_{z}=(\tilde x_{1z},...,\tilde x_{Nz})$, and $\tilde Y_z\in \mathcal{D}_Y$ with $\mathcal{D}_y=\{-1<z<1,\,\,\,-\sqrt{1-z^2}<\tilde y_{z}<\sqrt{1-z^2}\}$. The same value of $z$ is used to define all components of $\tilde x_{z}$ and $\tilde y_z$, so we can define the domain of all points in the spherical extension as:

\begin{equation}
\mathcal{D}=\{ (\tilde x_z,\tilde y_z,z),\,\,\,-1<z<1,\,\,\,-\sqrt{1-z^2)}<\tilde x_{nz}<\sqrt{1-z^2},\,\,\,-\sqrt{1-z^2}<\tilde y_{z}<\sqrt{1-z^2}\}.
\nonumber
\end{equation}

We are now ready to introduce the generalization of the path integral that we aimed for:

\begin{definition}[Geometric Path Integral]
Let $\gamma(t)$, $0\leq t\leq 1$ be a path in $\mathcal{D}$ with $\gamma(0)=(0,0,-1)$ and $\gamma(1)=(0,0,1)$. Let $\Gamma$ be the set of all such paths, and let $D\gamma$ be a suitable measure on $\Gamma$. We define the geometric path integral associated to condition \eqref{classH}, and dependent on a parameter $h>0$, as:

\begin{equation}
\label{NPI}
P(a,m,h)=\int_{\gamma \in \Gamma} e^{\frac{\imath}{h}\int_{0}^1 L(\gamma(t),a,m)dt}D\gamma
\end{equation}
\end{definition}
In general the integrand $L$ would not be defined at the extremes of the paths $\gamma$. But since we chose it to be the square of a (linear transformation of a) neural network classifier $F$ with exponential sigmoidal activation functions, then for each path $\gamma$ the integral is well defined as $z \rightarrow \pm 1$, since the exponential flattening of the sigmoidal functions (\cite{classbook2}, page 225) assure that the limits of $L$ for $z \rightarrow \pm 1$, with extension maps $\tilde x_{nz}=\tilde x_n \sqrt{1-z^2}$, and all other variables fixed, do actually exists.

\begin{remark}
The choice of the appropriate measure $D\gamma$ that ensures convergence of path integrals is very delicate and it will require further investigation in the context of geometric path integrals for appropriate classes of integrands $L$. Note, however, that the geometrical path integral is defined on compact sets, and this is a scenario where the standard path integrals are amenable to rigorous convergence results \cite{semiclassical}.
\end{remark}

In order to understand the motivation for the integral in \eqref{NPI}, we consider the case in which $L$ (and therefore $H$) has a set of zeros in $\mathcal{D}\bigcap \{(\tilde x_z,\tilde y_z,z):z=0\}$, specifically we can assume that $L(\tilde x_0, \tilde y_0, 0,a,m)=0$. We can then build a full path $\gamma$ in $\mathcal{D}$ such that $L(\gamma(t),a,m)=0$ for every $t$ in $[0,1]$, just by taking suitable mappings of $(\tilde x_0, \tilde y_0)$ in $\mathcal{D}$ for all values of $-1<z<1$.

Now, following standard heuristic arguments for semi-cassical approximations of path integrals \cite{feynman}, \cite{condensed} chapter 3, \cite{schul} we expect the following result to hold. We denote by $\Re f$ and $\Im f$ the real and imaginary part of $f$ respectively.

\begin{proposition}[Geometric Path Integral Real Dominance Conditions] If the network $\dot x=f(x,a)$ can generate states belonging to class $C_m$, then $\Re(P(a,m,h))>0$, $\Re(P(a,m,h))>>\Im(P(a,m,h))$ for all positive values $h$ sufficiently close to zero.
\end{proposition}

\noindent
{\it Heuristic Justification:} Since $L=H^2$ in \eqref{L} is a quadratic functions, and  $H$ is linear in $y$, all the first derivatives of $L$ vanish only when $L$ itself is zero. Moreover, if $\dot x=f(x,a)$ can generate states belonging to class $C_m$, from Definition 2 we know that $F(X,f(X,a))-Y=0$ has a solution in the domain that establishes class belonging, and therefore $H=0$ has an appropriate solution as well (see \eqref{H} and \eqref{classH}). This implies that there is a path $\gamma_0\in \Gamma$ such that $L$ is identically zero on $\gamma_0$, and that the functional $S(\gamma,a,m)=\int_{0}^1 L(\gamma(t),a,m)dt$ has first order functional derivatives equal to zero as well, at $\gamma=\gamma_0$. The path $\gamma_0$ is therefore an extremal path for $S(\gamma,a,m)$.
In the limit of $h\rightarrow 0$, the extremal paths, and quadratic fluctuations around them, will dominate the geometric path integral, since all other non-extremal contributions to $P(a,m,h)$ will mostly cancel each other out because of the much faster phase  interference of the corresponding exponential integrals in $P(a,m,h)$. For near-extremal paths in a neighborhood of extremal paths, we have $e^{\frac{\imath}{h}\int_{0}^1 L(\gamma(t),a,m)dt}\approx e^{\frac{\imath}{h}\int_{0}^1 0dt}\approx 1$, and they will provide a large, real positive contribution to $P(a,m,h)$, so that $P(a,m,h)\approx p+\imath q$ with $p$ positive and $p>>q$, if $h$ is sufficiently small, possibly up to a multiplicative phase factor that depends only on quadratic local fluctuations of near-extremal paths around the paths for which $S(\gamma,a,m)=0$ \cite{semiclassical}.

\begin{remark}
In standard path integrals, there may be a different change of phase for each of the individual contributions of extremal paths to the overall integral (\cite{schul} chapter 17). Essentially, this is due to the fact that extremal paths may not be globally minima of the functionals that replace $S(\gamma,a,m)$ in standard path integrals. No such problem arises for the extremal paths used in our heuristic justification, since they all achieve the very minimum (zero) value allowed for the functional $S(\gamma,a,m)$ itself.
\end{remark}

\begin{remark}
For geometric path integrals, the extremal paths are not isolated, when they exist. This may require techniques from functional field integrals (i.e. higher dimensional path integrals, see \cite{condensed}, chapter 4) for the actual derivation of semi-classical types of approximations in the limit $h\rightarrow 0$.
\end{remark}

The computation of $P(a,m,h)$ is a global approach to identify zeros of functions in a specific real domain. Indeed there is a dependence on the original domain of biologically meaningful conditions that is hidden in the definition of the function $L$. However, we really want to know whether these zeros are persistent in a full measure subset $\tilde A$ of a domain $A$ of parameters. Because of this additional requirement, we need one more step before we can fully express the condition \eqref{classA} with the geometric path integral formalism. This is achieved by taking an ordinary integral of a function of $P(a,m,h)$ with respect to the parameter vector $a$, in the domain $A$ where we want to enforce robustness as in \eqref{classA}.

\begin{definition}[Robustness Function] The robustness function $R(m,h)$ associated to class $C_m$ is, for $h>0$:
\begin{equation}
\label{R}
R(m,h)=\int_A P(a,m,h)e^{-\frac{1}{h}(\Im{P(a,m,h)})^2}da
\end{equation}
\end{definition}

This definition of robustness may formally remind the reader of the one proposed by Kitano in \cite{kitano}. The two proposals, however, are substantially different, since Kitano considers a space of perturbations and defines a measure of robustness through integration on that space.

What is crucial for our interpretation of the CS problem is the fact that $R(m,h)$ inherits the real dominance conditions from $P(a,m,h)$, namely:

\begin{proposition}[ Robustness Function Real Dominance Conditions] If a phenotype class $C_m$ is robust for a region of parameters $\tilde A\subseteq A$, then the robustness function $R(m,h)$ satisfies the real dominance conditions, i.e., $\Re(R(m,h))>>\Im(R(m,h))$, and $\Re(R(m,h))>0$ for all positive values $h$ sufficiently close to zero.
\end{proposition}

\noindent
{\it Heuristic Justification:} We make the assumption that the imaginary part of $P(a,m,h)$ goes to zero fast enough as $h\rightarrow 0$ if
$P(a,m,h)$ satisfies the real dominant conditions, more particularly we assume $|\Im P(a,m,h)|\approx h^{1/2+\epsilon}$ with $\epsilon>0$ for $h$ small.
Now, from Definition 3, if a phenotype class $C_m$ is robust, then there are persistent zeros of the function $F(X,f(X,a))-Y=0$ in the appropriate domain, and $P(a,m,h)$ satisfies $\Re(P(a,m,h))>>\Im(P(a,m,h))$ and $\Re(P(a,m,h))>0$ for all $a$ in some region $\tilde A\subseteq A$. Therefore for $a$ in such region $\tilde A$, $P(a,m,h)$ will give large, real positive contributions to $R(m,h)$ for $h$ that goes to zero, since the exponential in \eqref{R} will converge to $1$.
Suppose instead we are in a region  $\bar A \subseteq A$ where the functional $S(\gamma,a,m)=\int_{0}^1 L(\gamma(t),a,m)dt$ has no extremal paths for all $a\in \bar A$.
Note that, reverting to a coordinate representation for $L$, for every small $\Delta a$, $L(\tilde x_z, \tilde y_z,z,a,m)$ and $L(\tilde x_z, \tilde y_z,z,a+\Delta a,m)$ will be equal on at most a finite number of points in $\mathcal{D}$, since we asked that $f(x,a)$ was analytical in $a$, and $F$ is also assumed analytical in its arguments. The differences, small, but located almost everywhere in $\mathcal{D}$, between $L(\tilde x_z, \tilde y_z,z,a,m)$ and $L(\tilde x_z, \tilde y_z,z,a+\Delta a,m)$ will be enhanced in the limit of $h\rightarrow 0$ leading to large differences in the phases of $P(a,m,h)$ and $P(a+\Delta a,m,h)$. Therefore nearby geometric path integrals in $\bar A \subseteq A$  will have uncorrelated phases for $h$ that goes to $0$. In particular, for each $h$, the set of points in $\bar A$ for which $\Im P(a,m,h)=0$ is of measure zero, and therefore this set can be removed when computing the integral in \eqref{R}. For all remaining $a\in \bar A$, the exponential in \eqref{R} will suppress to $0$ the contribution of the corresponding  $P(a,m,h)$ to $R(m,h)$ in the limit $h\rightarrow0$. We can conclude that the contributions to $R(m,h)$ from path integrals in $\bar A$ will be subject to strong phase interference, and also that their individual contributions to $R(m,h)$ will have norm that converges to $0$.
Putting together this result with the real dominant contributions from regions $\tilde A$ of $A$ for which $S(\gamma,a,m)$ has extremal paths, we conclude that  $R(m,h)$ will satisfy the real dominant conditions for all $h$ sufficiently close to $0$.

\begin{remark}
While the real dominant conditions of propositions 1 and 2 are only necessary conditions to the existence of zeros and persistent zeros for $H$ respectively, these conditions are likely to be sufficient for a generic $H$. In the absence of extremal paths for $S(\gamma,a,m)$, it is unlikely that real dominant conditions would hold for all $h$ sufficiently close to zero, as in the limit the phase of $P(a,m,h)$ becomes increasingly uncorrelated as a function of both $h$ and $a$. 
Also, if $e^{-\frac{1}{h}(\Im{P(a,m,h)})^2}$ in \eqref{R} is substituted by
$e^{-\frac{1}{h}(\frac{\Im{P(a,m,h)}}{\Re{P(a,m,h)}})^2}$, the condition 
$|\Im(P(a,m,h))|\approx h^{1/2+\epsilon}$ in the justification of Proposition 2 could be substituted by the weaker $|\Im(P(a,m,h))/\Re(P(a,m,h))|\approx h^{1/2+\epsilon}$, at the price of a slightly more complicated argument.
\end{remark}

\begin{remark}
The real dominance conditions for geometric path integrals seem to portend a method to establish the existence of solutions of equations (in particular real equations) in bounded domains, that does not depend on constraints on the signs of first derivatives.
More specifically, suppose that we want to know whether $g(x)=0$ has zeros in a domain $D$, then we can substitute the function $H$ in \eqref{classH} with $H(x,y)=(g(x)-y)^2$ (we have no dependence from $a$ and $m$ in this setting, and no change of variables). The boundaries for $x$ can be inferred directly from $D$ and we take $y$ in the domain $D_y(\epsilon)=\{ y: -\epsilon<y<\epsilon\}$ for $\epsilon>0$. We can use this function $H(x,y)$ in the definition of the geometric path integral, so that the partial derivatives of the corresponding function $L$ in \eqref{NPI} are all zeros only when $g(x)-y=0$. Therefore, if $g(x)$ has zeros in the domain $D$, then $g(x)-y=0$ at least for some $(x,y)\in D\times D_y(\epsilon)$, and  
a real dominance condition on the geometric path integral will hold on $D\times D_y(\epsilon)$ for all $\epsilon$ sufficiently small. We will explore this important application of our technique in a subsequent paper.
\end{remark}

\section{Centipedes' Segmentation Problem Reinterpreted}

We come back now to the problem that motivated this work. How to interpret a pattern of allowed changes in phenotype on the basis of the structure of the underlying molecular network? In the context of the centipedes' segmentation problem, we assume that the network $\dot x=f(x,a)$ is essentially the same for all species of centipedes, with only the set of parameters $a$ changing from one species to the other. This assumption is not unreasonable, if we think that the same species of centipedes can display different individuals with different number of segments, showing that there is, in the same network the potential for variable segmentations. Moreover, we would expect the process of segmentation to be evolutionary stable (\cite{minelli}, page 53).

Classify now the networks in such a way that the classification function mirrors the quantitative phenotype structure. In the setting of the CS problem, if $\dot x=f(x,a)$ gives rise to a phenotype with 15 pairs of legs, we assume that $15-1<F(X,\dot X)<15$ for all states $(X,\dot X)$ arising from that network. Effectively, we treat $F$ as a nonlinear regression model, which predicts the number of pairs of legs from state variables. Except for the fact that we do not simply want to know what is the output of $F$ under a specific input $(X,\dot X)$, but we must assure that some suitable $(X,\dot X)$ can be generated stably from the network $\dot x=f(x,a)$. We use the robustness function $R$ to formulate a quantitative version of the CS problem as follows:

\begin{problem}[CS Problem Reinterpreted] Let $\dot x=f(x,a)$ be an analytic network, polynomial in $x$, with $a\in A$ and $x\in D$ that describes the molecular dynamics of relevant signaling compounds in centipedes' embryos. Show that there is no interval classifier $F$, with growth conditions compatible with the geometric path integral definition in \eqref{NPI}, and trained on a set of actual data for known centipedes segmentation classes $C_{m_1},\dots,C_{m_k}$, such that, for all $h$ sufficiently close to zero, $R(m,h)$ satisfies the real dominant conditions for $m$ even, $m=17,19$.
\end{problem}

We assume that the integral defining $R$ is taken over a very large domain $A$, so that we can suppose that different segmentation phenotypes correspond to different regions of parameters within $A$.

According to the network path integral formalism, Problem 2 is equivalent to stating that it is not possible to find sizable volumes of parameters in $A$ such that  $\dot x=f(x,a)$ always gives rise to state variables $(X,\dot X)$ such that $F(X,\dot X)$ is even or equal to $17,19$; when $F(X,\dot X)$ is trained to properly predict the allowed, known number of pairs of legs of centipedes. The quantitative interpretation of the CS problem is not dependent on a specific classifier, and it rather enforces some properties on any classifier that we may derive from experimental data.

We defined interval classifiers in Section 2 exactly to be able to achieve this compact interpretation of the CS problem, the function $R(m,h)$ is linked to the corresponding class $C_m$ just by the single parameter $m$, that, in principle, could be treated now as a continuous variable.

\begin{remark}
The geometric path integral formulation of the CS problem allows us to comment on some essential differences between mathematization on biology and in physics. The theoretical tools used to solve problems can be similar in the two fields, as we suggested with the development of the geometric path integral formalism. But in the biological setting we lack the ability to unify our understanding of multiple problems: the functionals in the geometric path integral interpretation of the CS problem are derived from the classifiers found with supervised learning, and therefore they are not amenable to interpretation. It is as if every question that we may ask about phenotypes requires its own theory, and associated geometric path integral, not reducible, even in principle, to simpler geometric path integrals.
\end{remark}

\section{Challenges Ahead}

In this section we highlight some of the major problems that need to be addressed regarding geometric path integrals and their applications\footnote{ Refer to Remarks 5 an 6 for outstanding issues related to the definition of geometric path integrals.}.

First, it is to be seen how known analytical techniques to evaluate and approximate path integrals \cite{kleinert} apply to the highly non-standard geometrical path integrals derived from biological classification problems. At the very least, we would expect numerical approximation of these integrals to be possible, and hopefully less computationally intensive than an actual resolution of the associated geometrical problem in \eqref{classA}, especially for very large domains. 

Moreover, path and functional field integrals are powerful qualitative tools to describe the global state of large systems \cite{condensed}, and similar methods for geometric path  integrals may allow to rule out real dominant conditions for entire families of functions $H$. In particular, to approach the geometric path integral interpretation of the CS problem, we would need methods that can constrain effectively the sign of the real part of $R(m,h)$ for large spaces of classifiers trained on a set of experimental data. It would also be important to develop the theory of geometric path integrals to allow for a precise estimate of the size of the parameters for which \eqref{classA} is satisfied. This parameter size can vary for different classes, and therefore a careful estimate could be used to compare the relative robustness of different classes.

We point out that a network may be constrained by several classifier functions, if different phenotype characteristics are dependent on it. The function $L$ in the geometric path integral \eqref{NPI} can be extended to these cases by taking a sum of squares of the classifier functions where each of them requires the introduction of a new slack variable, and all heuristic arguments that lead to the real dominance conditions can be repeated in this generalized case as well.

Finally, we note that if the classifier function $F$ is fixed, it is possible to ask questions on the topology of the networks $\dot x=f(x,a)$ that are compatible with the real dominance conditions for each specific class. In other words, the analytical structure of suitable geometric path integrals may encode and shed light on the structure of the topologies of molecular networks that are compatible with some given phenotypic outcomes.

\end{document}